\documentclass[12pt]{article}
\usepackage{amsfonts}

\usepackage{amsmath}
\usepackage{graphicx}
\csname @addtoreset\endcsname{equation}{section}
\newcommand{\eq}{\begin{equation}}
\newcommand{\feq}{\end{equation}}
\newcommand{\eqn}{\begin{eqnarray}}
\newcommand{\feqn}{\end{eqnarray}}
\newcommand{\arr}{\begin{eqnarray*}}
\newcommand{\farr}{\end{eqnarray*}}

\font\mybb=msbm10 at 12pt
\def\bb#1{\hbox{\mybb#1}}

\def\ph{\varphi}

\newcommand{\VEV}[1]{\left\langle #1 \right\rangle}

\begin{document}

\begin{titlepage}
\begin{flushright}
IFUM-690-FT\\
hep-th/0106247
\end{flushright}
\vspace{.3cm}
\begin{center}
\renewcommand{\thefootnote}{\fnsymbol{footnote}}
{\Large \bf Some Aspects of the
de~Sitter/CFT Correspondence}
\vfill%\vskip 15mm%27.mm
{\large \bf {Dietmar Klemm\footnote{dietmar.klemm@mi.infn.it}}}\\
\renewcommand{\thefootnote}{\arabic{footnote}}
\setcounter{footnote}{0}
\vfill%\vskip 7mm%1cm
{\small
Dipartimento di Fisica dell'Universit\`a di Milano
and\\ INFN, Sezione di Milano,
Via Celoria 16,
20133 Milano, Italy.\\}
\end{center}
\vfill
\begin{center}
{\bf Abstract}
\end{center}
{\small We discuss several aspects of the proposed correspondence
between quantum gravity on de~Sitter spaces and Euclidean conformal
field theories. The central charge appearing in the asymptotic symmetry
algebra of three-dimensional de~Sitter space is derived both from
the conformal anomaly and the transformation law of the CFT stress tensor
when going from dS$_3$ in planar coordinates to dS$_3$ with cosmological
horizon. The two-point correlator for CFT operators coupling to bulk
scalars is obtained in static coordinates, corresponding to a CFT
on a cylinder. Correlation functions are also computed for CFTs
on two-dimensional hyperbolic space.
We furthermore determine the energy momentum tensor and the Casimir
energy of the conformal field theory
dual to the Schwarz\-schild-de~Sitter solution in five dimensions.
Requiring the pressure to be positive yields an upper bound for the
black hole mass, given by the mass of the Nariai solution. Beyond
that bound, which is similar to the one found by Strominger requiring
the conformal weights of CFT operators to be real, one encounters
naked singularities.}

\end{titlepage}

\section{Introduction}

In the last months there has been an increasing interest in gravity
on de~Sitter (dS) spacetimes \cite{Banks:2000fe,Balasubramanian:2001rb,
Witten:2001kn,Volovich:2001rt,Chamblin:2001dx,Strominger:2001pn,
Mazur:2001aa,Li:2001ky,Nojiri:2001mf}.
This is partially motivated by recent
astrophysical data indicating a positive cosmological
constant \cite{Perlmutter:2000rr}.
Apart from phenomenological aspects, it would also be desirable to
understand the role of de~Sitter spaces in string theory, and to
clarify the microscopic origin of the entropy of dS
space\footnote{For microscopic derivations of dS entropy based on the
Chern-Simons formulation of 2+1 dimensional dS gravity, or on
other approaches that are not directly related to string theory,
cf.~\cite{Maldacena:1998ih}-\cite{Hawking:2001da}.}. Whereas
string theory on anti-de~Sitter spaces is known to have a dual description
in terms of certain superconformal field theories \cite{Aharony:2000ti},
no such explicit duality was known up to now for dS spacetimes. The first
evidence for a dS/CFT correspondence was given by Hull \cite{Hull:1998vg},
who considered so-called
IIA* and IIB* string theories, which are obtained by T-duality on a
timelike circle from the IIB and IIA theories respectively. The type IIB*
theory admits E4-branes, which are the images of D4-branes under
T-duality along the time coordinate of the brane. The E4-branes interpolate
between Minkowski space at infinity and dS$_5$ $\times$ $H^5$ near the
horizon, where $H^5$ denotes hyperbolic space. The effective action
describing E4-brane excitations is
a Euclidean $D=4$, ${\cal N}=4$ $U(N)$ super Yang-Mills theory,
which is obtained from SYM in ten dimensions by reduction on a six-torus
with one timelike circle. This leads to a duality between type IIB*
string theory on dS$_5$ $\times$ $H^5$
and the mentioned euclidean SYM theory \cite{Hull:1998vg}. Unfortunately
this example is pathological, because both theories have
ghosts\footnote{Note however that Euclidean super Yang-Mills theory
can be twisted to obtain a well-defined topological field theory in
which the physical
states are the BRST cohomology
classes \cite{Acharya:1997gp,Acharya:1998jn,Blau:1997pp}.
According to \cite{Hull:1998vg}, this should correspond to a twisting
of the type IIB* string theory, with a topological gravity
limit.}.\\
Based on \cite{Hull:1998vg} and related ideas that appeared
in \cite{Hull:2000mt,Bousso:1999cb,Balasubramanian:2001rb},
Strominger proposed recently a more general holographic duality
relating quantum gravity on dS$_D$ to a conformal field theory residing
on the past boundary ${\cal I}^-$ of dS$_D$ \cite{Strominger:2001pn}.
He argued that in general
this CFT may be non-unitary, with operators having complex conformal weights,
if the dual bulk fields are sufficiently massive. The asymptotic
symmetry algebra of three-dimensional de~Sitter space was found to
consist of two copies of Virasoro algebras with central charges
$c = \tilde c = 3l/2G$ \cite{Strominger:2001pn},
where $l$ is the dS$_3$ curvature radius, and
$G$ denotes Newton's constant. This generalizes the result of Brown and
Henneaux \cite{Brown:1986nw} to the case of positive cosmological constant.
In this paper, we derive this central charge by two independent
alternate methods.
The first uses the relation to the trace anomaly, whereas the second
is based on the transformation law of the CFT stress tensor when going
from the plane, corresponding to dS$_3$ in planar coordinates, to
the cylinder, corresponding to dS$_3$ with a cosmological horizon.
The shift of $-c/24$ in the Virasoro generators is thereby identified
with the negative mass of dS$_3$ in presence of a cosmological horizon.\\
The two-point correlator for CFT operators coupling to bulk
scalars is then obtained in static coordinates. This correlation
function agrees with the result that one would get
by starting with the two-point function on the plane, and then using the
scaling relations for CFT operators under the coordinate transformation
from the plane to the cylinder. Thereby, the conformal weights of the
operators are given in terms of the masses of the bulk fields.
Several properties of the correlator in static coordinates are
discussed.\\
We also compute correlation functions for CFTs on two-dimensional
hyperbolic space, dual to dS$_3$ in hyperbolic slicing.\\
Finally we consider the Schwarz\-schild-dS black hole in arbitrary dimension,
and derive a Smarr-like formula. For the five-dimensional case,
we determine then the stress tensor and the Casimir energy
of the dual CFT. Requiring the pressure
to be positive yields an upper bound on the black hole mass, much like
the bound obtained in \cite{Strominger:2001pn} for bulk scalars, following from
reality of the CFT conformal weights. Our bound is exactly the mass
of the Nariai solution \cite{Nariai:1951}, for which the event horizon and
the cosmological horizon coalesce.

\section{The Stress Tensor for de~Sitter Space}

The gravitational action of $(n+1)$-dimensional de~Sitter gravity
has the form
\begin{equation}
I_{\mathrm{bulk}} + I_{\mathrm{surf}} =
    \frac{1}{16\pi G}\int_{{\cal M}}d^{n+1}x\sqrt{-g}\left(R -
    \frac{n(n-1)}{l^2}\right) + \frac{1}{8\pi G}\int_{\partial {\cal M}}d^nx
    \sqrt{\gamma} K\,.
    \label{action}
\end{equation}

The first term is the Einstein-Hilbert action with positive cosmological
constant $\Lambda = n(n-1)/2l^2$, whereas the second is the Gibbons-Hawking
boundary term necessary to have a well-defined variational principle.
$K$ is the trace of the extrinsic curvature $K_{\mu\nu} = -\nabla_{\left(
\mu\right.}n_{\left.\nu\right)}$ of the spacetime boundary $\partial {\cal M}$,
with $n^{\mu}$ denoting the outward pointing unit normal.
$\gamma$ is the induced metric on the boundary.
In evaluating expressions like (\ref{action}) one usually
encounters divergences coming
from integration over the infinite volume of spacetime. In the case
of adS gravity, a regularization procedure was proposed
in \cite{Balasubramanian:1999re,Mann:1999pc,Emparan:1999pm},
that consists of adding counterterms constructed from local curvature
invariants of the boundary. These counterterms, which are essentially unique,
can be easily generalized to the case of positive cosmological constant,
yielding
\begin{equation}
I_{\mathrm{ct}} = \frac{1}{8\pi G}\int_{\partial {\cal M}}d^nx \sqrt{\gamma}
         \left[\frac{n-1}{l} - \frac{l{\cal R}}{2(n-2)}\right],
         \label{counter}
\end{equation}

where ${\cal R}$ is the Ricci scalar of the boundary metric $\gamma$.
Using $I = I_{\mathrm{bulk}} + I_{\mathrm{surf}} + I_{\mathrm{ct}}$,
one can then construct a conserved stress tensor \cite{Brown:1993br}
associated to the boundary $\partial {\cal M}$,
\begin{eqnarray}
T^{\mu\nu} &=& -\frac{2}{\sqrt{\gamma}}\frac{\delta I}{\delta \gamma_{\mu\nu}}
               \nonumber \\
           &=& \frac{1}{8\pi G}\left[K^{\mu\nu} - K\gamma^{\mu\nu} -
               \frac{(n-1)}{l}\gamma^{\mu\nu} -
               \frac{l{\cal G}^{\mu\nu}}{(n-2)}\right], \label{stresst}
\end{eqnarray}
where ${\cal G}^{\mu\nu}$ is the Einstein tensor of $\gamma$.
This generalizes the result of \cite{Strominger:2001pn} for 
three-dimensional de~Sitter gravity, where the last term in
(\ref{counter}) and (\ref{stresst}) has to be omitted.

\section{dS$_3$ Central Charge from Conformal Anomaly}

We start from dS$_3$ in spherical slicing,

\begin{equation}
ds^2 = -l^2d\tau^2 + l^2\cosh^2\tau (d\theta^2 + \sin^2\theta d\phi^2)\,.
       \label{spheredS3}
\end{equation}

These coordinates cover the entire spacetime. The Carter-Penrose diagram
of (\ref{spheredS3}) can be found in \cite{Strominger:2001pn}.
Using (\ref{stresst}), we can now compute the stress tensor of the
dual Euclidean CFT, residing on the past boundary ${\cal I}^-$ ($\tau
\to -\infty$), i.~e.~ on a
two-sphere $S^2$. This yields

\begin{equation}
T_{\theta\theta} = -\frac{l}{16\pi G}\,, \qquad T_{\phi\phi} =
-\frac{l}{16\pi G}\sin^2\theta\,.
\end{equation}

The CFT metric $h_{\mu\nu}$ can be obtained by setting $d\tau=0$ in
(\ref{spheredS3}) and dropping the diverging conformal factor $\cosh^2\tau$.
This gives the trace anomaly

\begin{equation}
T = h^{\mu\nu}T_{\mu\nu} = -\frac{1}{8\pi Gl}\,.
\end{equation}

Comparing this with $T = -cR/24\pi$, where $R = 2/l^2$, we learn
that

\begin{equation}
c = \frac{3l}{2G}\,, \label{central}
\end{equation}

confirming the result of \cite{Strominger:2001pn}.\\
Alternatively, one could consider dS$_3$ in the coordinates
(cf.~appendix)

\begin{equation}
ds^2 = -l^2d\tau^2 + l^2\sinh^2\tau (d\theta^2 + \sinh^2\theta d\phi^2)\,,
       \label{hyperbolic}
\end{equation}

corresponding to a conformal field theory on the hyperbolic space $H^2$.
Then one gets

\begin{equation}
T_{\theta\theta} = \frac{l}{16\pi G}\,, \qquad T_{\phi\phi} =
\frac{l}{16\pi G}\sinh^2\theta\,,
\end{equation}

so that $T = 1/8\pi Gl$. As the scalar curvature is
now $R = -2/l^2$, this yields again (\ref{central}).\\
We finally note that the slicing (\ref{hyperbolic}) allows furthermore to
consider CFTs on compact Riemann surfaces of genus $g > 1$, by quotienting
dS$_3$ by a suitable discrete subgroup $\Gamma$ of the isometry group $SO(2,1)$
of $H^2$.

\section{Central Charge from Casimir Energy}

There is still another way to obtain the central charge (\ref{central}).
Consider dS$_3$ in planar coordinates, with metric

\begin{equation}
ds^2 = -d\tau^2 + e^{-2\tau/l}dzd\bar z\,, \label{planedS3}
\end{equation}

where $-\infty < \tau < \infty$, and $z,\bar z$ range over an infinite plane.
Using $n=-\partial_{\tau}$, it can be easily checked that the energy
momentum tensor (\ref{stresst}) vanishes for (\ref{planedS3}).\\
By means of the transformation
\begin{eqnarray}
\tau &=& t - \frac l2 \ln |V(r)|\,, \qquad V(r) = 1 - \frac{r^2}{l^2}\,,
         \nonumber \\
z &=& r|V(r)|^{-\frac 12}e^{-i\phi + t/l}\,, \label{transf} \\
\bar z &=& r|V(r)|^{-\frac 12}e^{i\phi + t/l}\,, \nonumber
\end{eqnarray}

\begin{figure}[ht]
\begin{center}
\includegraphics[width=0.4\textwidth]{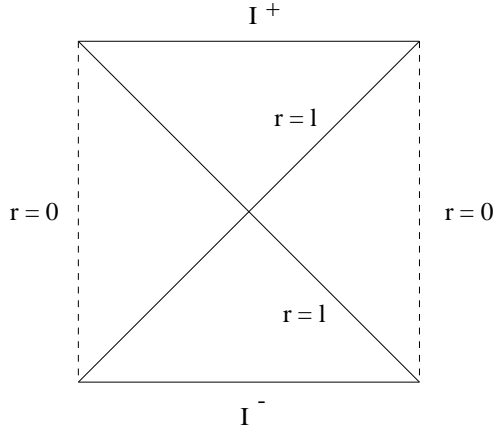}
\end{center}
\caption{\small{Carter-Penrose diagram for de~Sitter
space \cite{Gibbons:1977mu}. ${\cal I}^{\pm}$ correspond to $r = \infty$.}
}
\label{penrose}
\end{figure}

(\ref{planedS3}) can be cast into the form
\begin{equation}
ds^2 = -V(r)dt^2 + V(r)^{-1}dr^2 + r^2 d\phi^2\,, \label{cosmdS3}
\end{equation}

where $\phi$ is identified modulo $2\pi$.
The Carter-Penrose diagram of (\ref{cosmdS3}) is shown in figure
(\ref{penrose}). The past and future boundaries ${\cal I}^{\mp}$
correspond to $r = \infty$.

The characteristic feature is the appearance of a cosmological event
horizon \cite{Gibbons:1977mu} at $r=l$, with temperature $T = 1/2\pi l$
and Bekenstein-Hawking entropy $S = \pi l/2G$.
For $r>l$, the Killing vector
$\partial_t = l^{-1}(z \partial_z + \bar z \partial_{\bar z}) +
\partial_{\tau}$ becomes spacelike, whereas $\partial_r$ becomes timelike.
On the past boundary ${\cal I}^-$, we have $r \to \infty$, and
the transformation (\ref{transf}) becomes

\begin{equation}
z = le^{-i\phi + t/l}\,, \qquad
\bar z = le^{i\phi + t/l}\,,
\end{equation}

or equivalently $z = l\exp(-iw/l)$, where we defined
$w = l\phi + it$. This is precisely the
transformation from the plane to the cylinder, and it is well-known
that this induces a shift in the Virasoro generators of a two-dimensional
conformal field theory. The Hamiltonian of time translation in the $w$
frame is

\begin{equation}
lH = L_0 + \tilde{L}_0 - \frac{c + \tilde c}{24}\,.
\end{equation}

As the stress tensor (\ref{stresst}) vanishes for the metric (\ref{planedS3}),
corresponding to $z,\bar z$,
we know that the conformal weights $L_0$ and $\tilde{L}_0$ are zero.
On the other hand, we can compute the stress tensor for dS$_3$ in
static coordinates (\ref{cosmdS3}) (corresponding to $w,\bar w$),
and from this the Hamiltonian $H$.
A straightforward calculation yields for $r \to \infty$

\begin{equation}
T_{tt} = -\frac{1}{16\pi Gl}\,, \qquad T_{\phi\phi} = \frac{l}{16\pi G}\,.
         \label{stresscosm3d}
\end{equation}

The conserved charge $M$ associated with the Killing field $k=\partial_t$ is
given by \cite{Brown:1993br,Balasubramanian:1999re}

\begin{equation}
M = \int_0^{2\pi}T_{\mu\nu}u^{\mu}k^{\nu}\sqrt{\sigma}d\phi\,, \label{mass}
\end{equation}

where $u = (r^2/l^2 - 1)^{-1/2}\partial_t$ is the unit normal to the
surface $\Sigma_t$ of constant $t$ in $\partial {\cal M}$, and $\sigma$
denotes the induced metric on $\Sigma_t$. One then gets

\begin{equation}
M = -\frac{1}{8G}\,.
\end{equation}

Equating this with the Hamiltonian $H$, and using $c = \tilde c$, we
finally obtain

\begin{equation}
c = \frac{3l}{2G}\,.
\end{equation}

\section{Two-point Correlators in Static Coordinates}

Like in \cite{Strominger:2001pn}, we can now compute correlation
functions of CFT operators that couple to bulk fields, in the spirit
of the AdS/CFT correspondence. For simplicity, we will consider only
massive scalar fields. In the static coordinates (\ref{cosmdS3}),
the Klein-Gordon equation reads

\begin{equation}
m^2\Phi = \nabla^2\Phi = \frac 1r \partial_r (rV(r)\partial_r \Phi)
          - \frac 1V \partial_t^2 \Phi + \frac{1}{r^2}\partial_{\phi}^2 \Phi\,.
          \label{KGcyl}
\end{equation}

Near ${\cal I}^-$, the last two terms in (\ref{KGcyl}) can be neglected,
leading to the asymptotic behaviour

\begin{equation}
\Phi \sim r^{-h_{\pm}}
\end{equation}

for $r \to \infty$, with $h_{\pm}$ given by

\begin{equation}
h_{\pm} = 1 \pm \sqrt{1 - m^2l^2}\,.
\end{equation}

We will consider only the case $0 < m^2l^2 <1$, which implies real
$h_{\pm}$. Similar to \cite{Strominger:2001pn}, we impose the boundary
condition

\begin{equation}
\lim_{r\to\infty} \Phi(r,t,\ph) = r^{-h_-}\Phi_-(t,\phi)\,. \label{asymptPhi}
\end{equation}

The two-point function of the operator ${\cal O}$ coupling to $\Phi$
can be obtained \cite{Strominger:2001pn} from

\begin{equation}
\lim_{r\to\infty}\int_{{\cal I}^-} dt\, d\phi\, dt'\, d\phi'\,
\left[\frac{(rr')^2}{l^2}\left(\Phi(r,t,\phi)
{\buildrel \leftrightarrow \over \partial_{r_{\ast}}}G(r,t,\phi; r',t',\phi')
{\buildrel \leftrightarrow \over \partial_{r'_{\ast}}}\Phi(r',t',\phi')
\right)\right]_{r=r'}, \label{expr}
\end{equation}

where $G$ denotes the Hadamard two-point function given
in (\ref{hadamard}), and $dr_{\ast} = (-V(r))^{-1/2}dr$.
For $r\to\infty$, $G$ behaves like (cf.~(\ref{asympthada}))

\begin{eqnarray}
\lefteqn{\lim_{r,r'\to\infty}G(r,t,\phi; r',t',\phi') = } \nonumber \\
&& \gamma_+ (rr')^{-h_+}\left[\cosh\frac{\Delta t}{l} - \cos\Delta\phi
\right]^{-h_+} + \gamma_- (rr')^{-h_-}\left[\cosh\frac{\Delta t}{l} -
\cos\Delta\phi\right]^{-h_-}\!\!, \label{asymptG}
\end{eqnarray}

where $\gamma_{\pm}$ are constants, and $\Delta t = t-t'$,
$\Delta \phi = \phi - \phi'$. Using (\ref{asymptPhi})
and (\ref{asymptG}), the expression (\ref{expr}) reduces modulo
a constant prefactor to

\begin{equation}
\int_{{\cal I}^-}dt\,d\phi\,dt'\,d\phi'\,\Phi_-(t,\phi)\frac{\mathrm{const.}}
{\left[\cosh\frac{\Delta t}{l} - \cos\Delta\phi\right]^{h_+}}
\Phi_-(t',\phi')\,.
\end{equation}

This yields the two-point correlator

\begin{equation}
\VEV{{\cal O}(t,\phi){\cal O}(t',\phi')} = \frac{\mathrm{const.}}
{\left[\cosh\frac{\Delta t}{l} - \cos\Delta\phi\right]^{h_+}}\,,
\label{2ptfcttphi}
\end{equation}

or in the coordinates $w,\bar w$,

\begin{equation}
\VEV{{\cal O}(w,\bar{w}){\cal O}(w',\bar{w}')} =
\mathrm{const.}\frac{e^{-\pi i Th_+(\Delta w + \Delta \bar{w})}}
{\left(1 - e^{-2\pi i T \Delta w}\right)^{h_+}
\left(1 - e^{-2\pi i T \Delta \bar{w}}\right)^{h_+}}\,, \label{2ptfct}
\end{equation}

where $T=1/2\pi l$ denotes the Hawking temperature of the cosmological
horizon. The AdS analogue of (\ref{2ptfct}) has been obtained in
\cite{Keski-Vakkuri:1999nw}. In that case, it corresponds to a BTZ black
hole in the bulk. (\ref{2ptfct}) agrees with the result that one would get
by starting with the two-point correlator

\begin{equation}
\VEV{{\cal O}(z,\bar{z}){\cal O}(z',\bar{z}')} \sim
\frac{1}{(\Delta z \Delta \bar{z})^{h_+}}\,, \label{corrplane}
\end{equation}

obtained in \cite{Strominger:2001pn}, and using the scaling relations
for dimension $(h_+,h_+)$ operators under the coordinate transformation
from the plane to the cylinder \cite{Cardy:1984}.\\
Note that the finite extent of the system is not in $t$ direction,
but along the coordinate $\phi$. The infinite cylinder geometry that
we have should thus correspond rather to a quantum chain at zero temperature,
but with periodic boundary conditions, and not to an infinite chain at
finite temperature. The system would thus have a correlation length
$\xi = l/h_+$, and a mass gap $\delta E = h_+/l$ between the ground state
and the first excited state \cite{DiFranc}. Alternatively, one could
interpret (\ref{2ptfct}) as a thermal correlator at imaginary temperature
$iT = i/2\pi l$,

\begin{equation}
\VEV{{\cal O}(w,\bar{w}){\cal O}(w',\bar{w}')} \sim
\left[\sinh(i\pi T\Delta w)\sinh(i\pi T\Delta\bar w)\right]^{-h_+}\,.
\end{equation}

This would mean that the coordinate transformation $z = l\exp(-iw/l)$
induces a Bogoljubov transformation of the operators ${\cal O}(z,\bar z)$
to new operators ${\cal O}(w,\bar w)$ that see the Poincar\'{e} vacuum
as a thermal bath of excitations (at imaginary temperature).
Note also that (\ref{2ptfcttphi}) is invariant under
the shift

\begin{equation}
\Delta t \mapsto \Delta t + i\beta\,,
\end{equation}

where $\beta = 1/T$. This means that the temperature is imaginary,
because $t$ is already a Euclidean time. The significance of an imaginary
temperature in this context remains rather obscure, but it may be related
to the fact that ${\cal I}^-$ lies actually behind the horizon,
whereas the concept of a (real) Hawking temperature is well-defined
only outside the horizon.\\
It is possible to rederive the central charge (\ref{central}) by
expanding (\ref{2ptfcttphi}) in powers of $\Delta t$ and
$\Delta \phi$\footnote{I would like to thank Tassos Petkou
for pointing out this to me.},
which yields the leading term (\ref{corrplane}) plus finite size
corrections. Inserting the correct prefactor for the two-point function
of dimension $(h_+,h_+)$ operators on the cylinder \cite{Cardy:1987dg}, we
have

\begin{equation}
\VEV{{\cal O}(t,\phi){\cal O}(0,0)} = l^{-2h_+}
\left[2\cosh\frac{t}{l} - 2\cos\phi\right]^{-h_+}\,,
\end{equation}

which yields for small $t, \phi$

\begin{equation}
\VEV{{\cal O}(t,\phi){\cal O}(0,0)} = (t^2 + l^2\phi^2)^{-h_+}
\left[1 - \frac{h_+}{12l^2}(t^2 - l^2\phi^2)
+ \ldots\right]\,.
\end{equation}

One can now compare this with the operator product
expansion \cite{Cardy:1987dg}

\begin{equation}
{\cal O}({\mathbf{r}}){\cal O}(0,0) \sim r^{-2h_+} + C^{\mu\nu}({\mathbf{r}})
T_{\mu\nu}(0) + \ldots\,,
\end{equation}

where in our conventions

\begin{equation}
C^{\mu\nu}({\mathbf{r}}) = \frac{2\pi h_+}{c}\left(r^{\mu}r^{\nu} - \frac 12
r^2 h^{\mu\nu}\right)r^{-2h_+}\,.
\end{equation}

Using the expectation value (\ref{stresscosm3d}) of the energy momentum tensor,
one obtains again $c=3l/2G$.

\section{Two-point Correlators in Hyperbolic Coordinates}

In \cite{Strominger:2001pn}, the two-point function of CFT operators
that couple to massive Klein-Gordon bulk fields was computed for
the case where the CFT lives on a sphere $S^2$. Starting from dS$_3$
in hyperbolic coordinates (\ref{hyperbolic}), we can do similar
calculations for CFTs on $H^2$. The Klein-Gordon equation reads

\begin{eqnarray}
\lefteqn{m^2 l^2 \Phi = l^2 \nabla^2\Phi = } \nonumber \\
&& - \partial_{\tau}^2 \Phi - 2\coth\tau \partial_{\tau}
\Phi + \frac{1}{\sinh^2\tau \sinh\theta}\partial_{\theta}(\sinh\theta
\partial_{\theta}\Phi) + \frac{1}{\sinh^2\tau \sinh^2\theta}
\partial_{\phi}^2\Phi\,. \label{KGhyper}
\end{eqnarray}

For $\tau \to -\infty$, the last two terms in (\ref{KGhyper})
can be neglected, and thus

\begin{equation}
\lim_{\tau \to -\infty}\Phi(\tau,\theta,\phi) = \Phi_+(\theta,\phi)
e^{h_+\tau} + \Phi_-(\theta,\phi)e^{h_-\tau}\,. \label{asPhihyp}
\end{equation}

The asymptotic behaviour of the propagator is (cf.~(\ref{asympthadahyp}))

\begin{eqnarray}
\lefteqn{\lim_{\tau, \tau' \to \infty}G(\tau, w, \bar w; \tau', v, \bar v) = }
\nonumber \\
&& \gamma_+ e^{h_+(\tau + \tau')}\frac{(1 - w \bar w)^{h_+}
(1 - v \bar v)^{h_+}}{|w - v|^{2h_+}} +
\gamma_- e^{h_-(\tau + \tau')}\frac{(1 - w \bar w)^{h_-}(1 - v \bar v)^{h_-}}
{|w - v|^{2h_-}}\,, \label{asGhyp}
\end{eqnarray}

where $w = \tanh\frac{\theta}{2}e^{i\phi}$ is a complex coordinate
on $H^2$. The hyperbolic analogue of (\ref{expr}) reads

\begin{equation}
\lim_{\tau \to -\infty}\int_{{\cal I}^-} d^2w d^2v \sqrt{h(w)h(v)}
\left[e^{-2(\tau + \tau')}\left(\Phi(r,w,\bar w)
{\buildrel \leftrightarrow \over \partial_{\tau}}G(\tau,w,\bar w;
\tau',v,\bar v)
{\buildrel \leftrightarrow \over \partial_{\tau'}}\Phi(\tau',v,\bar v)
\right)\right]_{\tau=\tau'}, \label{exprhyp}
\end{equation}

$\sqrt{h(w)} = 2(1 - w\bar w)^{-2}$ denoting the measure on $H^2$.
Inserting (\ref{asPhihyp}) and (\ref{asGhyp}) into (\ref{exprhyp}),
one obtains, up to a normalization constant,

\begin{equation}
\int_{{\cal I}^-} d^2w\, d^2v\, \sqrt{h(w)h(v)}\,
(\gamma_+ \Phi_-(w, \bar w)\Delta_{h_+}\Phi_-(v, \bar v) +
\gamma_- \Phi_+(w, \bar w)\Delta_{h_-}\Phi_+(v, \bar v))\,,
\end{equation}

where $\Delta_{h_{\pm}}$ is the two-point correlator for a conformal
field of dimension $h_{\pm}$ on the hyperbolic space $H^2$,

\begin{equation}
\Delta_{h_{\pm}} = \left[\frac{(1 - w\bar w)(1 - v\bar v)}{|w - v|^2}
\right]^{h_{\pm}}\,. \label{2pthyp}
\end{equation}

\section{Schwarzschild-de~Sitter Black Holes in Arbitrary Dimension}

The metric for the Schwarz\-schild-de~Sitter black hole in $D$ dimensions
reads\footnote{For generalizations cf.~\cite{Kastor:1993nn}
-\cite{Klemm:2001gh}.}

\begin{equation}
ds^2 = -V(r) dt^2 + V(r)^{-1}dr^2 + r^2 d\Omega_{D-2}^2, \label{BHdSD}
\end{equation}

where

\begin{equation}
V(r) = 1 - \frac{\mu}{r^{D-3}} - \frac{r^2}{l^2},
\end{equation}

and $d\Omega_{D-2}^2$ denotes the standard metric on the unit $S^{D-2}$.
For $D=5$, the product of (\ref{BHdSD}) with hyperbolic space $H^5$
is a solution of IIB* supergravity \cite{Hull:1998vg}, whereas for
$D=4$, the de~Sitter black hole (\ref{BHdSD}) times adS$_7$ solves the
equations of motion \cite{Hull:2000mt} of the low energy
effective action of M* theory \cite{Hull:1998ym},
which is the strong coupling limit of the IIA* theory, and has signature
$9+2$.\\
For mass parameters $\mu$ with $0 < \mu < \mu_N$, where

\begin{equation}
\mu_N = \frac{2l^{D-3}}{D-1}\left(\frac{D-3}{D-1}\right)^{\frac{D-3}{2}}\,,
\end{equation}

one has a black hole in
de~Sitter space with event horizon at $r=r_H>0$ and cosmological horizon
at $r=r_C > r_H$, with $V(r_H) = V(r_C) = 0$. For $\mu = \mu_N$,
the event horizon and the cosmological horizon coalesce, and one
gets the Nariai solution \cite{Nariai:1951}. For $\mu > \mu_N$,
(\ref{BHdSD}) describes a naked singularity in de~Sitter space.
We thus notice that the absence of naked singularities yields
an upper bound for the black hole mass. Below we will see
(for the case $D=5$) that the same
bound results from the requirement that the pressure of the dual CFT
must be positive. It is interesting that analogous bounds arise for
bulk fields coupling to CFT operators, if one wants to have real
conformal weights \cite{Strominger:2001pn}.\\
We start by deriving a Smarr-like formula for (\ref{BHdSD}),
following the lines of \cite{Gibbons:1977mu}.
Consider the Killing identity

\begin{equation}
\nabla_{\mu}\nabla_{\nu}k^{\mu} = R_{\nu\rho}k^{\rho} =
\frac{D-1}{l^2}k_{\nu}\,, \label{killingid}
\end{equation}

where $k^{\mu}$ is a Killing vector, $\nabla_{\left(\nu\right.}k_{\left.\mu
\right)} = 0$, and we used the Einstein equations in the last step.
Now integrate (\ref{killingid}) on a spacelike hypersurface
$\Sigma_t$ from the black hole horizon $r_H$ to the cosmological horizon
$r_C$. On using Gauss' law, this gives

\begin{equation}
\frac 12 \int_{\partial \Sigma_t} \nabla_{\mu}k_{\nu} d\Sigma^{\mu\nu}
= \frac{D-1}{l^2} \int_{\Sigma_t} k_{\nu} d\Sigma^{\nu}\,,
  \label{killingidint}
\end{equation}

where the boundary $\partial \Sigma_t$ consists of the intersection
of $\Sigma_t$ with the black hole- and the cosmological horizon,

\begin{equation}
\partial \Sigma_t = S^{D-2}(r_H) \cup S^{D-2}(r_C)\,.
\end{equation}

Applying (\ref{killingidint}) to the Killing vector $k = \partial_t$ yields

\begin{eqnarray}
\frac{D-2}{16\pi G (D-3)} \int_{S^{D-2}(r_H)} \nabla_{\mu}k_{\nu}
d\Sigma^{\mu\nu} &+&
\frac{D-2}{16\pi G (D-3)} \int_{S^{D-2}(r_C)} \nabla_{\mu}k_{\nu}
d\Sigma^{\mu\nu} \nonumber \\
&=&\frac{(D-1)(D-2)}{8\pi G l^2 (D-3)} \int_{\Sigma_t} k_{\nu} d\Sigma^{\nu}\,,
\label{massformula1}
\end{eqnarray}

where $G = l_P^{D-2}$ denotes Newton's constant. One can regard the right-hand
side of Eq.~(\ref{massformula1}) as representing the contribution of
the cosmological constant to the mass within the cosmological horizon.
As in \cite{Gibbons:1977mu}, we therefore identify the second term on
the left-hand side as the total mass $M_C$ within the cosmological
horizon, and the first term as the negative of the black hole mass
$M_{BH}$. The latter can be rewritten by using the definition of the
surface gravity $\kappa_H$, which, by the zeroth law, is constant on the
horizon. In this way, one obtains

\begin{equation}
M_{BH} = \frac{D-2}{8\pi G (D-3)}\kappa_H A_H\,,
\end{equation}

where $A_H$ denotes the area of the event horizon. One therefore gets
the Smarr-type formula

\begin{equation}
M_C = \frac{D-2}{8\pi G (D-3)}\kappa_H A_H +
\frac{(D-1)(D-2)}{8\pi G l^2 (D-3)} \int_{\Sigma_t} k_{\nu} d\Sigma^{\nu}\,,
\label{smarr}
\end{equation}

from which a first law of black hole mechanics can be derived.
Evaluating (\ref{smarr}) yields

\begin{equation}
M_C = \frac{(D-2)V_{D-2}}{16\pi G}\left[\mu -
\frac{2}{D-3}\frac{r_C^{D-1}}{l^2}\right]\,, \label{MC}
\end{equation}

where $V_{D-2}$ is the volume of the unit $S^{D-2}$.\\
Alternatively, we can use the
stress tensor (\ref{stresst}) in order to associate a mass to the
spacetime (\ref{BHdSD}). We will consider only the case $D=5$, and
write the metric on the unit $S^3$ as

\begin{equation}
d\Omega_3^2 = d\theta^2 + \sin^2\theta d\phi^2 + \cos^2\theta d\psi^2\,.
\end{equation}

Then the stress tensor (\ref{stresst}) of the CFT dual to the
Schwarz\-schild-de~Sitter black hole in five dimensions reads

\begin{eqnarray}
8\pi G T_{tt} &=& \frac{12\mu - 3l^2}{8lr^2} + {\cal O}(r^{-4})\,,
                  \nonumber \\
8\pi G T_{\theta\theta} &=& \frac{-4\mu l + l^3}{8r^2} + {\cal O}(r^{-4})\,,
                  \label{stresst5d} \\
8\pi G T_{\phi\phi} &=& \frac{-4\mu l + l^3}{8r^2}\sin^2\theta +
                  {\cal O}(r^{-4})\,, \nonumber \\
8\pi G T_{\psi\psi} &=& \frac{-4\mu l + l^3}{8r^2}\cos^2\theta +
                  {\cal O}(r^{-4})\,. \nonumber \\
\end{eqnarray}

Like in the AdS/CFT correspondence, the metric of the manifold on which the
CFT resides, is defined by

\begin{equation}
h_{\mu\nu} = \lim_{r\to\infty} \frac{l^2}{r^2}\gamma_{\mu\nu},
\end{equation}

where $\gamma_{\mu\nu}$ denotes the induced metric on the boundary
$\partial {\cal M}$. The field theory's stress tensor $\hat{T}^{\mu\nu}$
is related to the one in (\ref{stresst}) by the rescaling
\cite{Myers:1999qn}

\begin{equation}
\sqrt{h}h_{\mu\rho}\hat{T}^{\rho\nu} = \lim_{r\to\infty}\sqrt{\gamma}
\gamma_{\mu\rho}T^{\rho\nu}\,,
\end{equation}

which amounts to multiplying all expressions for $T_{\mu\nu}$ in
(\ref{stresst5d}) by $r^2/l^2$ before taking the limit $r\to\infty$.
Defining the "pressure"

\begin{equation}
p = \frac{-4\mu + l^2}{64\pi Gl^3}\,,
\end{equation}

and $v = (1,0,0,0)$,
we can write the energy-momentum tensor as

\begin{equation}
\hat{T}_{\mu\nu} = p(-4v_{\mu}v_{\nu} + h_{\mu\nu})\,,
\end{equation}

which is conserved and traceless. Note that the pressure is positive
precisely if $\mu < \mu_N$, i.~e.~if naked singularities are absent.\\
The conserved charge associated with the Killing vector $k=\partial_t$
is given by

\begin{equation}
M = \int T_{\mu\nu}u^{\mu}k^{\nu}\sqrt{\sigma}d\theta d\phi d\psi\,,
\end{equation}

where, as in (\ref{mass}), $u = (-V(r))^{-1/2}\partial_t$ is the
unit normal to the surface $\Sigma_t$ of constant $t$ in $\partial {\cal M}$,
and $\sigma$ denotes the induced metric on $\Sigma_t$.
One then gets

\begin{equation}
M = \frac{3\pi}{8G}\left(\mu - \frac{l^2}{4}\right)\,. \label{M}
\end{equation}

This is to be identified with the energy of the dual CFT. Note that
$M$ is negative for $\mu < \mu_N$. The Casimir energy $E_C$ of the conformal
field theory, which lives on $\bb{R} \times S^3$, is obtained
by setting $\mu = 0$, i.~e.~, for pure de~Sitter space without a black
hole. We have thus

\begin{equation}
E_C = -\frac{3\pi l^2}{32 G}\,.
      \label{casimir}
\end{equation}

It is interesting to note that this can be obtained from the mass of
anti-de~Sitter space, $E_C = 3\pi l^2/32G$ \cite{Balasubramanian:1999re},
corresponding to the
Casimir energy of ${\cal N}=4$ super Yang-Mills theory on
$\bb{R} \times S^3$, by simply replacing $l^2 \to -l^2$.\\
We also notice that
$M$ in (\ref{M}) is different from $M_C$ (\ref{MC}), because the latter
keeps track only of
the mass within the cosmological horizon, whereas the former
has contributions also from outside.

\section*{Note added}

Some time after this paper was posted on the web, a related
computation of the Schwarz\-schild-de~Sitter mass (\ref{M})
appeared in \cite{Balasubramanian:2001nb}. There the same result
(\ref{M}) was obtained, but with opposite overall sign. This sign
difference comes from the definition

\begin{equation}
T^{\mu\nu} = + \frac{2}{\sqrt{\gamma}} \frac{\delta I}{\delta \gamma_{\mu\nu}}
\end{equation}

for the stress tensor of the dual CFT used in \cite{Balasubramanian:2001nb},
whereas in (\ref{stresst})
we defined $T^{\mu\nu} = - (2/\sqrt{\gamma}) \delta I/\delta
\gamma_{\mu\nu}$\footnote{The definition (\ref{stresst}) was also used in
\cite{Strominger:2001pn}.}. The motivation for choosing the opposite
sign in \cite{Balasubramanian:2001nb} was to get an energy that increases
with the entropy. Indeed the entropy of the cosmological horizon,

\begin{equation}
S_C = \frac{r_C^3\pi^2}{2G}\,,
\end{equation}

increases with decreasing $\mu$ (for $\mu = 0$, it reaches its maximum
value $S_C = l^3\pi^2/2G$). At the same time, the mass

\begin{equation}
M = \frac{3\pi}{8G}\left(\frac{l^2}{4} - \mu\right)\,, \label{Mbala}
\end{equation}

obtained in \cite{Balasubramanian:2001nb} rises if we lower $\mu$.
The temperature $T_C$ of the cosmological horizon, derived from the
thermodynamic fundamental relation as $T_C = \partial M/\partial S$, is
then positive. Things change however if we consider the entropy $S_H$
of the black hole event horizon instead of the entropy of the
cosmological horizon. Inserting $\mu = r_{H/C}^2 - r_{H/C}^4/l^2$ and
$S_{H/C} = r_{H/C}^3\pi^2/2G$ into (\ref{M}), we obtain, up to the
Casimir term (\ref{casimir}), the fundamental relation

\begin{equation}
M(S) = \frac{3\pi}{8l_P}\left[\left(\frac{2S}{\pi^2}\right)^{2/3} -
       \frac{l_P^2}{l^2}\left(\frac{2S}{\pi^2}\right)^{4/3}\right]\,,
\end{equation}

where $S$ can be either $S_H$ or $S_C$. The temperature is then given by

\begin{equation}
T_{H/C} = \frac{\partial M}{\partial S_{H/C}} = \frac{1}{2\pi}\left(
          \frac{1}{r_{H/C}} - \frac{2r_{H/C}}{l^2}\right)\,.
\end{equation}

As we have $r_H^2 \le l^2/2$ and $r_C^2 \ge l^2/2$, the black hole temperature
is positive with our definition, whereas the temperature associated to
the cosmological horizon is negative. If we use instead the mass
(\ref{Mbala}) of \cite{Balasubramanian:2001nb}, $T_C$ is positive, whereas
the Hawking temperature $T_H$ of the black hole is negative, or, in
other words, the mass decreases with increasing black hole entropy.

\section*{Acknowledgements}
\small

This work was partially supported by INFN, MURST and
by the European Commission RTN program
HPRN-CT-2000-00131, in which D.~K.~is associated to the University of
Torino. The author would like to thank S.~Cacciatori and A.~C.~Petkou for
useful discussions, and A.~Strominger for clarifying correspondence.
\normalsize

\begin{appendix}

\section{Coordinate Systems in de~Sitter Space}

\subsection{Static Coordinates}

Consider $\bb{R}_1^{D+1}$ with coordinates $X^A$, $A = 0, 1, \ldots, D$,
and metric $(\eta_{AB}) = \mathrm{diag}(-1, 1, \ldots, 1)$.
$D$-dimensional
de~Sitter space dS$_D$ can then be defined as the hypersurface

\begin{equation}
\eta_{AB}X^A X^B = l^2\,. \label{hyper}
\end{equation}

Fix now

\begin{equation}
(X^0)^2 - (X^D)^2 = -l^2\left(1 - \frac{r^2}{l^2}\right) = -l^2V(r)\,.
\label{hyperbola}
\end{equation}

One has then

\begin{equation}
(X^1)^2 + \ldots + (X^{D-1})^2 = r^2\,,
\end{equation}

so the coordinates $X^1, \ldots, X^{D-1}$ range over a $(D-2)$-sphere
$S^{D-2}$ with radius $r$. If we parametrize the hyperbola (\ref{hyperbola})
of fixed $r$ by

\begin{equation}
X^0 = \sqrt{r^2-l^2}\cosh\frac tl\,, \qquad
X^D = \sqrt{r^2-l^2}\sinh\frac tl\,,
\end{equation}

then the induced metric on the hypersurface (\ref{hyper}) is given by

\begin{equation}
ds^2 = -V(r)dt^2 + V(r)^{-1}dr^2 + r^2 d\Omega_{D-2}^2\,, \label{dScosmo}
\end{equation}

where $d\Omega_{D-2}^2$ denotes the standard metric on the unit $S^{D-2}$.
(\ref{dScosmo}) describes de~Sitter space in static coordinates,
with horizon at $r=l$.\\

Below, we shall compute the de~Sitter invariant Hadamard two-point
function. To this aim, we need
the geodesic distance $d(X, X')$ between two points $X$ and $X'$ on
(\ref{hyper}), which reads \cite{Strominger:2001pn}

\begin{equation}
d = l\arccos P\,,
\end{equation}

where

\begin{equation}
l^2P(X, X') = X^A \eta_{AB} {X'}^B\,.
\end{equation}

In the coordinates (\ref{dScosmo}), one
gets

\begin{equation}
l^2P(X, X') = -\sqrt{r^2-l^2}\sqrt{{r'}^2-l^2}\cosh
           \frac{t-t'}{l} + rr'\cos\Theta\,, \label{Pcosmo}
\end{equation}

where $\Theta = \Theta(\Omega, \Omega')$ denotes the geodesic distance
of two points with angular variables $\Omega$ and $\Omega'$
on the unit $S^{D-2}$,
e.~g.~$\Theta = \phi-\phi'$ for $D=3$.\\

Alternatively, we can fix

\begin{equation}
(X^{D-1})^2 + (X^D)^2 = \tau^2 + l^2\,. \label{circle}
\end{equation}

This yields

\begin{equation}
-(X^0)^2 + (X^1)^2 + \ldots + (X^{D-2})^2 = -\tau^2\,,
\end{equation}

so the coordinates $X^0, X^1, \ldots, X^{D-2}$ range over a hyperbolic
space $H^{D-2}$ with curvature radius $\tau$. Parametrize the circle
(\ref{circle}) of fixed $\tau$ by

\begin{equation}
X^{D-1} = \sqrt{\tau^2 + l^2}\cos\frac{v}{l}\,, \qquad
X^D = \sqrt{\tau^2 + l^2}\sin\frac{v}{l}\,.
\end{equation}

This leads to the induced metric

\begin{equation}
ds^2 = -\left(1 + \frac{\tau^2}{l^2}\right)^{-1}d\tau^2 +
\left(1 + \frac{\tau^2}{l^2}\right)dv^2 +
\tau^2d\Sigma_{D-2}^2
\end{equation}

on the hypersurface (\ref{hyper}). $d\Sigma_{D-2}^2$ is now the
standard line element on the unit hyperbolic space $H^{D-2}$.

\subsection{Hyperbolic Coordinates}

If we set instead

\begin{equation}
X^D = l\cosh\tau\,,
\end{equation}

then the coordinates $X^0, X^1, \ldots, X^{D-1}$ range over
$H^{D-1}$ with curvature radius $l\sinh\tau$. The induced metric
on (\ref{hyper}) takes the form

\begin{equation}
ds^2 = -l^2d\tau^2 + l^2\sinh^2\tau d\Sigma_{D-1}^2\,. \label{dShyp}
\end{equation}

Let us write the hyperbolic metric $d\Sigma_{D-1}^2$ as

\begin{equation}
d\Sigma_{D-1}^2 = d\theta^2 + \sinh^2\theta d\Omega_{D-2}^2\,,
\end{equation}

where $d\Omega_{D-2}^2$ is the line element on the unit $S^{D-2}$.
This yields

\begin{eqnarray}
P(X, X') &=& \cosh\tau\cosh\tau' - \sinh\tau\sinh\tau'\cosh\theta
\cosh\theta' \nonumber \\
&+& \sinh\tau\sinh\tau'\sinh\theta\sinh\theta'\cos\Theta\,, \label{Phyp}
\end{eqnarray}

where again $\Theta = \Theta(\Omega, \Omega')$ denotes the geodesic distance
of two points with angular variables $\Omega$ and $\Omega'$
on the unit $S^{D-2}$.

\section{Green's Functions}

The de~Sitter invariant Hadamard two-point function

\begin{equation}
G(X, X') = \mathrm{const.}\VEV{0|\{\Phi(X), \Phi(X')\}|0}
\end{equation}

obeys

\begin{equation}
(\nabla^2_x - m^2)G(x, x') = 0\,, 
\end{equation}

and depends on $x$ and $x'$ only through $P$, as de~Sitter space is
maximally symmetric. As shown in \cite{Candelas:1975du}, one can write

\begin{equation}
l^2(\nabla^2_x - m^2)f(P) = (1-P^2)\frac{d^2f}{dP^2} - DP\frac{df}{dP}
- m^2l^2f
\end{equation}

for an arbitrary function $f$. This leads to the differential equation

\begin{equation}
(P^2 - 1)\frac{d^2G}{dP^2} + DP\frac{dG}{dP} + m^2l^2G = 0
\end{equation}

for the Hadamard two-point function, with the solution

\begin{equation}
G(P) = \mathrm{Re} F(h_+, h-, \frac D2; \frac{1+P}{2})\,,
       \label{hadamard}
\end{equation}

where $F$ is a hypergeometric function, and

\begin{equation}
h_{\pm} = \frac 12(D-1 \pm \sqrt{(D-1)^2 - 4m^2l^2})\,.
\end{equation}

(\ref{hadamard}) generalizes the result of \cite{Strominger:2001pn}
to arbitrary dimension.\\
In the static coordinates (\ref{dScosmo}),
one has the asymptotic behaviour near ${\cal I}^-$

\begin{equation}
\lim_{r,r'\to\infty} P(r,t,\Omega; r',t',\Omega') = -\frac{rr'}{l^2}
\left[\cosh\frac{t-t'}{l} - \cos\Theta\right]\,,
\end{equation}

which diverges. Like in \cite{Strominger:2001pn}, we can then use the
formula

\begin{eqnarray}
\lefteqn{F(h_+, h_-, \frac D2; z) = } \label{formula} \\
&& \frac{\Gamma(\frac D2)\Gamma(h_- - h_+)}
{\Gamma(h_-)\Gamma(\frac D2 - h_+)}(-z)^{-h_+}F(h_+, h_+ - \frac D2 + 1,
h_+ - h_- + 1; \frac 1z) + (h_+ \leftrightarrow h_-)\,, \nonumber
\end{eqnarray}

together with $F(\alpha, \beta, \gamma; 0) = 1$, to obtain

\begin{eqnarray}
\lefteqn{\lim_{r,r'\to\infty} G(r,t,\Omega; r',t',\Omega') = } \nonumber \\
&& \!\!\!\!\frac{\Gamma(\frac D2)\Gamma(h_- - h_+)}
{\Gamma(h_-)\Gamma(\frac D2 - h_+)}\left(\frac{rr'}{2l^2}\right)^{-h_+}
\left[\cosh\frac{t-t'}{l} - \cos\Theta\right]^{-h_+} + (h_+
\leftrightarrow h_-)\,.
\label{asympthada}
\end{eqnarray}

The asymptotic scaling behaviour of the two-point function $G$ was
first observed in \cite{Antoniadis:1997fu}.\\

In the hyperbolic coordinates (\ref{dShyp}),
one has the asymptotic behaviour near ${\cal I}^-$

\begin{equation}
\lim_{\tau,\tau'\to-\infty} P(\tau,\theta,\Omega; \tau',\theta',\Omega') =
\frac 14 e^{-\tau-\tau'}\left[1 - \cosh\theta\cosh\theta' +
\sinh\theta\sinh\theta'\cos\Theta\right]\,,
\end{equation}

which again diverges. Using (\ref{formula}), we get

\begin{eqnarray}
\lefteqn{\lim_{\tau,\tau'\to\infty} G(\tau,\theta,\Omega; \tau',\theta',
\Omega') = } \nonumber \\
&& \!\!\!\!\frac{\Gamma(\frac D2)\Gamma(h_- - h_+)}
{\Gamma(h_-)\Gamma(\frac D2 - h_+)}8^{h_+}e^{h_+(\tau + \tau')}
\left[\cosh\theta\cosh\theta' - \sinh\theta\sinh\theta'\cos\Theta -
1\right]^{-h_+} \label{ashadahyp1} \\
&& + (h_+\leftrightarrow h_-)\,. \nonumber
\end{eqnarray}

For $D=3$, using the complex coordinate $w = \tanh\frac{\theta}{2}e^{i\phi}$,
(\ref{ashadahyp1}) reads

\begin{eqnarray}
\lefteqn{\lim_{\tau,\tau'\to\infty} G(\tau, w, \bar w; \tau', v,
\bar v) = } \nonumber \\
&& \!\!\!\!\frac{\Gamma(\frac 32)\Gamma(h_- - h_+)}
{\Gamma(h_-)\Gamma(\frac 32 - h_+)}4^{h_+}e^{h_+(\tau + \tau')}
\left[\frac{(1-w\bar w)(1-v\bar v)}{|w-v|^2}\right]^{h_+}
+ (h_+\leftrightarrow h_-)\,.
\label{asympthadahyp}
\end{eqnarray}

\end{appendix}

\end{document}